\documentclass[aps,floatfix,amsmath,amssymb,nofootinbib,twocolumn]{revtex4-1}
\usepackage{graphicx}% Include figure files
\usepackage{dcolumn}% Align table columns on decimal point
\usepackage{bbm}% bold math
\usepackage{epstopdf}
\usepackage[bookmarks=true,colorlinks,linkcolor=blue,urlcolor=blue,citecolor=blue]{hyperref}
\usepackage{bbold}
\usepackage{color}

\usepackage{setspace} %line spacing
\newcommand{\beq}{\begin{equation}}
\newcommand{\eeq}{\end{equation}}
\newcommand{\beqn}{\begin{eqnarray}}
\newcommand{\eeqn}{\end{eqnarray}}
\newcommand{\nn}{\nonumber\\}
\newcommand{\eulergamma}[1]{\Gamma \left( #1 \right)}

\usepackage{amsmath}
\usepackage{amssymb}

\def \q{{\mathbf{q}}}

\def \Q{{\mathbf{Q}}}
\def \G{{\mathbf{G}}}
\def \k{{\mathbf{k}}}
\def \K{{\mathbf{K}}}
\def \q{{\mathbf{q}}}
\def \P{{\mathbf{P}}}
\def \p{{\mathbf{p}}}

\def \tn{\textnormal}

\begin{document}

\title{Incommensurate $2k_F$ density wave quantum criticality in two dimensional metals}

\author{Johannes Halbinger}
\author{Dimitri Pimenov}
\author{Matthias Punk}
\affiliation{Physics Department, Arnold Sommerfeld Center for Theoretical Physics, Center for NanoScience, and Munich Center for Quantum Science and Technology (MCQST), Ludwig-Maximilians University Munich, Germany}

\date{\today}

\begin{abstract}

We revisit the problem of two dimensional metals in the vicinity of a quantum phase transition to incommensurate $\Q=2k_F$ charge density wave order, where the order parameter wave vector $\Q$ connects two hot spots on the Fermi surface with parallel tangents. Earlier theoretical works argued that such critical points are potentially unstable, if the Fermi surface at the hot spots is not sufficiently flat. Here we perform a controlled, perturbative renormalization group analysis and find a stable fixed point corresponding to a continuous quantum phase transition, which exhibits a strong dynamical nesting of the Fermi surface at the hot spots. We derive scaling forms of correlation functions at the critical point and discuss potential implications for experiments with transition metal dichalcogenides and rare-earth tellurides.

\end{abstract}

\maketitle

\section{Introduction}

While Landau's Fermi liquid theory has been tremendously successful in describing properties of ordinary metals, a variety of strongly correlated electron materials show an unusual strange metal or non-Fermi liquid behaviour, which is not captured within the Fermi liquid framework. It is usually characterized by a linear temperature dependence of the resistivity, as well as an absence of resistivity saturation at the Mott-Ioffe-Regel limit \cite{Gunnarson}, which is taken as evidence for the absence of well-defined electronic quasiparticle excitations \cite{EmeryKivelson}. Examples include the strange metal phases observed in cuprate and iron pnictide superconductors \cite{Hussey, Kasahara}, heavy fermion materials \cite{StewartRMP2001} and also in twisted bilayer graphene \cite{Cao2019}.

Quantum critical points in metals are a promising theoretical scenario giving rise to non-Fermi liquid phenomenology \cite{LohneysenReview}.
Indeed, in two spatial dimensions the strong coupling between gapless order parameter fluctuations and particle-hole excitations at the Fermi surface leads to a loss of electronic quasiparticle coherence and to a strong damping of order parameter fluctuations. Even though the computation of transport coefficients in these models remains a big challenge, two notable theoretical developments have considerably advanced our understanding of metallic quantum critical points: first, some lattice models of electrons coupled to a bosonic order parameter can be numerically studied using determinant quantum Monte-Carlo methods avoiding the infamous fermion sign problem \cite{BergScience, LedererSC, Gerlach, Bergreview}. Second, epsilon expansions have been developed for hot spot models of quantum critical points in metals, allowing to characterize non-Fermi liquid fixed points within a controlled renormalization group approach, where the bosonic and fermionic degrees of freedom are treated on equal footing \cite{Senthil, Dalidovich, Sur, Leereview}.

Due to the fact that strange metals are often found in the regime between a magnetically ordered phase and an ordinary Fermi liquid, a lot of theoretical work has focused on the study of commensurate spin density wave quantum criticality in metals \cite{Gerlach, Sur, Chubukov, Abanov, Metlitski, Meier, Patel, Maier, Schlief}.
In this work we consider incommensurate charge density wave (CDW) quantum critical points in quasi two dimensional metals instead, where the electron density spontaneously breaks translational symmetries and develops a density modulation with a wave vector $\Q$ that is incommensurate with the underlying crystalline lattice. In particular, we are interested in systems where the CDW ordering wave vector $\Q=2k_F$ is determined by a partial nesting condition of the Fermi surface and connects two points on the Fermi surface with parallel tangents. This is to be distinguished from perfect nesting, where entire sections of the Fermi surface are connected by the same $2k_F$ wavevector. 

The properties of $2k_F$ density wave quantum critical points in two-dimensional metals have been analyzed in previous theoretical works \cite{Altshuler, Bergeron, Holder, WangChubukov, Punk2015, Sykora, Jang2019}. While an early study by Altshuler \emph{et al.}~\cite{Altshuler} concluded that the incommensurate transition is of first order due to strong fluctuations, a more recent article by Sykora \emph{et al.}~\cite{Sykora} pointed out that the transition is potentially continuous, if the Fermi surface is sufficiently flat at the hot spots.
In this work, building upon the epsilon expansion by Dalidovich and Lee \cite{Dalidovich}, we resolve this open problem by performing a controlled renormalization group (RG) analysis of such an incommensurate $\Q=2k_F$ CDW transition. We show that there is a strong dynamical nesting of the Fermi surface at the two hot spots connected by the $2k_F$ wave vector and identify a stable RG fixed point corresponding to a continuous quantum phase transition. Furthermore, we compute critical exponents and the scaling form of correlation functions at the non-Fermi liquid fixed point to leading order in epsilon and point out experimental signatures.

CDW order plays an interesting role in underdoped cuprates and has been observed in a variety of quasi two-dimensional materials such as transition metal dichalcogenides and rare-earth tellurides. Several of these materials exhibit a CDW ordering wave vector which is commensurate with the crystalline lattice, implying that the transition is likely driven by the coupling to phonons. A few notable exceptions with incommensurate CDW order exist, such as the 2H forms of NbSe$_2$ and TaS$_2$ \cite{Wilson1975, Scholz1982}, VSe$_2$ \cite{Chen2018}, as well as SmTe$_3$ and TbTe$_3$ \cite{Gweon, Kapitulnik}. In some of these compounds the CDW transition temperature can be tuned to zero across a potential quantum critical point by intercalation or applying pressure, indicating that electronic correlations could be the main driving force behind the CDW transition \cite{Feng}. Moreover, some of these materials become superconducting at low temperatures in the vicinity of the putative CDW quantum critical point \cite{Feng2, Morosan, He}. One of our aims is to provide clear experimental signatures of an incommensurate $2k_F$ CDW quantum phase transition, which would allow to settle the question if the incommensurate CDW transition in materials like NbSe$_2$ is driven by electronic correlations, or by a different mechanism, such as the coupling to phonons.

The rest of this work is outlined as follows: in Sec.~\ref{sec2} we introduce the model of electrons coupled to charge density wave fluctuations in two dimensions, as well as a generalization to higher dimensions which is amenable to dimensional regularization. Sec.~\ref{sec3} presents our results for the one-loop boson and fermion self-energies in arbitrary dimensions. The RG flow equations, their fixed point structure and the scaling form of the boson and fermion two-point correlators are presented in Sec.~\ref{sec4}. A discussion of experimental signatures follows in Sec.~\ref{sec5}. Finally, results on superconducting instabilities in the vicinity of the QCP are presented in Sec.~\ref{sec6}. We close with discussions and conclusions in Sec.~\ref{sec7}.

\section{Model}
\label{sec2}

\begin{figure}
\begin{center}
\includegraphics[width=0.7 \columnwidth]{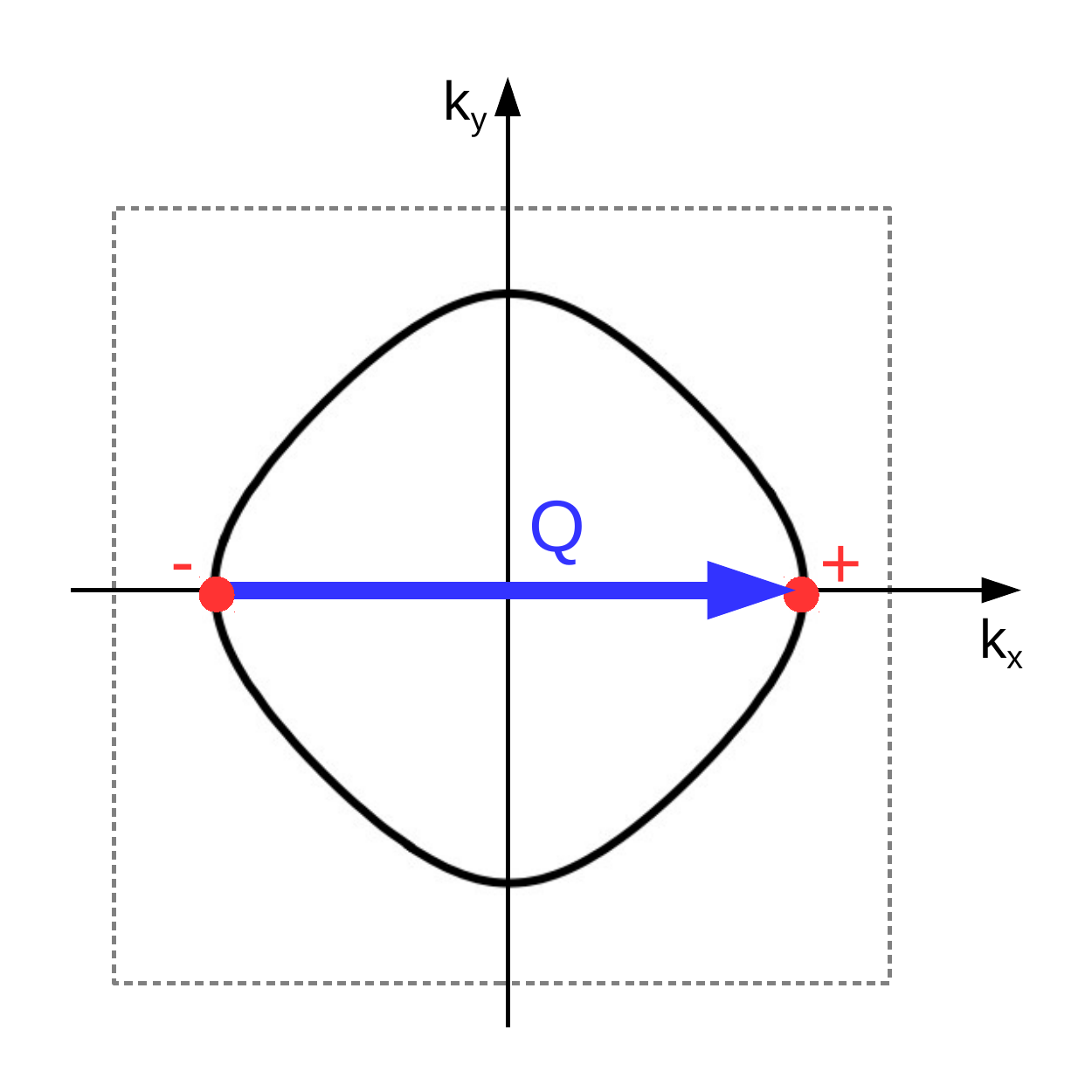}
\caption{Scattering geometry for electrons coupled to incommensurate $\Q=2k_F$ order parameter fluctuations. The order parameter wave vector $\Q$ (blue arrow) couples electrons in the vicinity of two hot spots (denoted by $-$ and $+$), where the Fermi surface (thick black line) has parallel tangents.}
\label{fig1}
\end{center}
\end{figure}

We start from a theory of electrons coupled to charge density wave fluctuations in two spatial dimensions described by the Euclidean action
	\beqn
 		S &=& \int_k \psi^\dagger(k) (-i k_0 +\xi_\k) \psi(k) + \frac{1}{2} \int_q \phi(q) \, \chi_q^{-1} \, \phi(-q) \notag \\
 		&& + \lambda \int_{k,q} \phi(q) \psi^\dagger(k+q) \psi(k) \ ,
	\eeqn
where the fermionic field $\psi(k)$ (spin index suppressed) describes electrons with frequency/momentum $k=(k_0,\k)$, the electron dispersion measured from the Fermi energy is denoted by $\xi_\k$, the real field $\phi(q)$ describes CDW fluctuations and $\chi_q=\chi_{-q}$ is the bare CDW susceptibility, which we assume to be peaked at the incommensurate $2k_F$ wave vectors $\pm\Q$. Consequently, electrons scatter predominantly in the vicinity of two hot-spots connected by the vector $\Q$ (see Fig.~\ref{fig1}). A finite order parameter  expectation value $\langle \phi \rangle \neq 0$ gives rise to a ground state with a spatially modulated electron density.  

We expand the action around the hot spots  by writing $\q = \pm \Q + \p$ and denoting $\phi(q_0,\pm\Q+\p) \equiv \phi^\pm(q_0,\p) \equiv  \phi^\pm(p)$. Analogously we denote the Fermion fields in the vicinity of the two hot spots by $\psi(k_0,\pm \Q/2+\k) \equiv \psi_{\pm}(k)$. Expanding the electron dispersion as well as the CDW susceptibility to second order around the hot spot momenta we thus obtain the low energy effective action
	\beqn
		S &=& \sum_{s=\pm} \sum_{j=1}^N \int_{k} \psi_{s,j}^{\dagger}(k) \left( -i k_0 + s k_x + k_y^2 \right) \psi_{s,j}(k) \nn
		 &+& \int_k \phi^+(k) \left( k_0^2 + k_x^2 + k_y^2 \right) \phi^-(-k) \nn
		  &+&\frac{\lambda}{\sqrt{N}} \sum_{j=1}^N \int_{k,p} \Big[ \phi^+(p) \psi^{\dagger}_{+,j}(k+p) \psi_{-,j}(k) \nn
		  &+& \phi^-(-p) \psi^{\dagger}_{-,j}(k-p) \psi_{+,j}(k) \Big],
		  \label{S0}
	\eeqn
where we've generalized the model to allow for $N$ distinct fermionic species (with $N=2$ for spin-1/2 fermions) and tuned the model to the quantum critical point, where the mass term for the order parameter field vanishes. Note that momenta in the fermion kinetic term are rescaled such that all proportionality constants are equal to unity. As discussed in detail below, only the $\sim k_y^2$ term in the kinetic part of the boson is relevant in the RG sense, while all other terms are irrelevant and will be discarded in the following. For this reason we have set their proportionality constants to unity as well. 

In order to perform an epsilon expansion we generalize this action to arbitrary dimensions by increasing the co-dimension of the Fermi surface, following earlier work by Dalidovich and Lee \cite{Dalidovich}. For this reason it is convenient to define the spinor
	\beq
		\Psi_j(k) = \begin{pmatrix} \psi_{+,j}(k) \\ \psi^{\dagger}_{-,j}(-k) \end{pmatrix},
	\eeq
and to rewrite the action \eqref{S0} as
	\beqn
		S &=& \sum_j \int_k \overline{\Psi}_j(k) \left[ - i \sigma_y k_0 + i \sigma_x \delta_k \right] \Psi_j(k) \nn
		 &+& \int_k \phi^+(k) \left( k_0^2 + k_x^2 + k_y^2 \right) \phi^-(-k) \notag \\
		 &-& \frac{i \lambda}{2 \sqrt{N}} \sum_j \int_{k,p} \Big[ \phi^+(p) \overline{\Psi}_j(k+p) \sigma_y \overline{\Psi}^T_j(-k)\nn
		 &+& \phi^-(-p) \Psi^T(p-k) \sigma_y \Psi(k) \Big]
	\eeqn
with
	\beq
		\overline{\Psi} = \Psi^{\dagger} \sigma_y,\ \delta_k=k_x + k_y^2.
	\eeq
A canonical way to generalize this action to $d$ spatial dimensions while keeping the action local is to write the fermion kinetic term as
	\beq
		\sum_j \int_{k^{d+1}} \overline{\Psi}_j(k) \left[ - i \bold{\Gamma} \cdot \K + i \sigma_x \delta_k \right] \Psi_j(k),
	\eeq
where we defined
	\beqn
		&&\bold{\Gamma} = \left( \sigma_y, \sigma_z, ..., \sigma_z \right),\ \K = \left( k_0, k_1, ..., k_{d-2} \right) = (k_0,\k),\nn
		&&\delta_k = k_{d-1} + k_d^2,\ \int_{k^{d+1}} = \int \frac{d^{d-1}\K dk_{d-1} dk_d}{(2\pi)^{d+1}} \ .
	\eeqn
The additional momenta $k_1, \dots , k_{d-2}$ correspond to the new directions perpendicular to the Fermi surface and due to the Dirac structure of the action the fermions have a linear dispersion in these directions. The kinetic term of the boson is generalized accordingly to 
	\beq
		\int_{k^{d+1}} \phi^+(k) \left( \K^2 + k_{d-1}^2 + k_d^2 \right) \phi^-(-k).
	\eeq
in arbitrary dimensions. \par
The quadratic terms in the action are invariant under the scaling transformation
	\beqn
		&&\K = \frac{\K'}{b},\ k_{d-1} = \frac{k_{d-1}'}{b},\ k_d = \frac{k_d'}{b^{\frac{1}{2}}}, \nn
		&&\Psi(k) = \Psi'(k') b^{\frac{d}{2} + \frac{3}{4}},\ \phi^{\pm}(k) = \phi'^{\pm}(k') b^{\frac{d}{2} + \frac{3}{4}}.
	\eeqn
A consequence of this scaling transformation is that all terms in the boson propagator apart from the $\sim k_d^2$ term are irrelevant at tree level and can be neglected in the following computations. However, as we will discuss in detail below, the renormalization group flow generates a \emph{linear} term $\sim k_{d-1}$ in the boson propagator, which is relevant, allowed by symmetry and crucial to remove infrared divergences. This term arises from the fact that the susceptibility is enhanced along the line of $2k_F$ wave vectors $\k$ defined via $\xi_{(\k+\G)/2}=0$ with $\G$ an arbitrary reciprocal lattice vector, connecting points on the Fermi surface with parallel tangents \cite{Holder}. For this reason we add the term $a k_{d-1}$ to the boson propagator from the start, where $a$ is a dimensionless parameter which will flow under the RG. 
The coupling constant $\lambda$ transforms as 
	\beq
		\lambda' = \lambda b^{\frac{1}{2} \left( \frac{5}{2}-d \right)},
	\eeq
consequently interactions are irrelevant in $d>5/2$ and we can perform a controlled expansion in small $\epsilon=\frac{5}{2} - d$. As usual we define a dimensionless coupling constant by introducing an arbitrary mass scale $\mu$ via the replacement $\lambda \rightarrow \lambda \mu^{\epsilon/2}$. Our final form of the action in general dimensions thus reads
	\beqn
		S &=& \sum_j \int_{k^{d+1}} \overline{\Psi}_j(k) \left[ - i \bold{\Gamma} \cdot \K + i \sigma_x \delta_k \right] \Psi_j(k) \nn
		 &+& \int_{k^{d+1}} \phi^+(k) \, \left( k_d^2 + a k_{d-1} \right) \, \phi^-(-k) \nn
		 &-& \frac{i \lambda \mu^{\epsilon/2}}{2\sqrt{N}} \sum_j \int_{k^{d+1},p^{d+1}} \Big[ \phi^+(p) \overline{\Psi}_j(k+p) \sigma_y \overline{\Psi}^T_j(-k) \nn
		 &+& \phi^-(-p) \Psi^T(p-k) \sigma_y \Psi(k) \Big].
		 \label{Sfinal}
	\eeqn
In the following we study this action within a field-theoretic renormalization group approach using dimensional regularization and the minimal subtraction scheme. For this reason we compute one-loop diagrams and extract the $1/\epsilon$ counterterms in the next section.

\section{One-loop diagrams}
\label{sec3}

The bare fermion and boson propagators for the theory in Eq.~\eqref{Sfinal} take the form
	\beqn
		G(k) &=& \left\langle \Psi(k) \overline{\Psi}(k) \right\rangle_0 = -i \frac{- \bold{\Gamma} \cdot \K + \sigma_x \delta_k}{\K^2 + \delta_k^2} \nn
		D_+(k) &=&  \left\langle \phi^+(k) \phi^-(-k) \right\rangle_0 = \frac{1}{k_d^2+a k_{d-1}} .
	\eeqn
Analogously we define $D_-(k) =  \left\langle \phi^-(k) \phi^+(-k) \right\rangle_0 \equiv D_+(-k)$. Even though this seems like a redundant definition, it is important to distinguish bosonic degrees of freedom in the vicinity of the $2k_F$ wave vector $\Q$ and $-\Q$ and the linear term $\sim a k_{d-1}$ in the boson propagator is allowed by this symmetry. The one-loop boson self-energy $\Pi_+(k) \equiv \Pi_-(-k)$ is given by the integral
	\beq
		\Pi_+(k)=-\frac{\lambda^2 \mu^{\epsilon}}{2 N} \int_{p^{d+1}} \text{Tr} \left[ \sigma_y G(p) \sigma_y G^T(k-p) \right] \label{bosonselfenergy}
	\eeq
and evaluates to (details can be found in appendix \ref{appA})
	\beq
		\Pi_+(k) = - u_1 \lambda^2 \frac{e_k}{\epsilon} - u_2 \lambda^2 \frac{\vert \K \vert^{\frac{3}{2}}}{\sqrt{\vert e_k \vert}} \Theta(-e_k) + \dots, \label{bosonpole}
	\eeq
where $\Theta(x)$ is the unit step function and we expanded the self-energy around $\epsilon \approx 0$ as well as around $\vert \K \vert \approx 0$ and defined 
	\beqn
		&&u_1 = \frac{\Gamma \left( \frac{5}{4} \right)}{8 \sqrt{2} \pi^{\frac{7}{4}}} \approx 0.0108,\ u_2 = \frac{\eulergamma{\frac{1}{4}}\eulergamma{\frac{5}{4}}}{16 \sqrt{2} \pi^{\frac{5}{4}}\eulergamma{\frac{7}{4}}} \approx 0.0378, \nn
		&&e_k = k_{d-1} + \frac{1}{2} k_d^2.
	\eeqn
Note that $e_k=0$ defines the line of $2k_F$ momenta. The frequency dependent term $\sim \vert \K \vert^{3/2}$ is the $d=5/2$ dimensional analog of the Landau damping term $\sim \vert k_0 \vert$ in two dimensions. Note that this term doesn't have a $1/\epsilon$ pole and thus does not renormalize. Consequently no frequency dependent terms are generated in the boson propagator during the one-loop RG flow. On the other hand, the $\sim 1/\epsilon$ term is proportional to $k_d^2$ and to $k_{d-1}$ and thus generates an RG flow of the corresponding terms in the boson propagator.

The fermion self-energy is given by the integral
	\beq
		\Sigma(k) = \frac{\lambda^2 \mu^{\epsilon}}{N} \int_{p^{d+1}} \sigma_y G^T(p-k) \sigma_y D_+(p). \label{fermionselfenergy}
	\eeq
As shown in appendix \ref{appB}, this integral evaluates to
%	\beqn
%		\Sigma(q)-\Sigma(0) = &-& \frac{2 u_1 \lambda^2 }{(1-a) \sqrt{\vert \tilde{a} \vert} N} \frac{i \sigma_x q_{d-1}}{\epsilon} \nn
%		&+& \frac{2 u_1 \lambda^2}{(1-a)^2 \sqrt{\vert \tilde{a} \vert} N} \frac{i \sigma_x q_{d}^2}{\epsilon} + \tn{finite terms}, \label{fermionpole}
%	\eeqn
	\beqn
		\Sigma(k)-\Sigma(0) &=& \frac{i \sigma_x}{\epsilon} \frac{2 u_1 \lambda^2}{(1-a) \sqrt{\vert \tilde{a} \vert} N} \left( \frac{k_d^2}{1-a} - k_{d-1}  \right) \nn
		&& + \ \tn{finite terms}, \label{fermionpole}
	\eeqn
where $\tilde{a} = \frac{a}{1-a}$. As discussed in detail below, these terms renormalize the fermion dispersion. The fact that all frequency dependent terms in the boson propagator are irrelevant implies that the fermionic self-energy has no frequency dependence either. This seems strange in the light of naive $1/N$ expansions, where the Landau damping term plays a prominent role and leads to a non-Fermi liquid form of the fermion self-energy. In any case, simple $1/N$ expansions are known to break down for models of metallic quantum critical points \cite{Metlitski, Lee2009, MetlitskiIsing}. Since no frequency dependent terms renormalize the boson propagator at one-loop level, Landau damping effects are less important within this controlled epsilon expansion scheme and only appear at two-loop order. Note that this is a crucial difference to the Ising-nematic QCP studied in Ref.~\cite{Dalidovich}, where the Landau damping term had to be included in the boson propagator to cure an IR divergence in the fermion self-energy, despite the fact that it doesn't renormalize at one loop order. By contrast, in the problem studied here an analogous IR divergence is cured by the $a k_{d-1}$ term in the boson propagator, as can be seen directly from the $a$ dependence in Eq.~\eqref{fermionpole}. 

Finally, we note that there is no one-loop vertex correction in our theory, because one simply cannot draw a one-loop vertex diagram given the structure of the interaction term in Eq.~\eqref{Sfinal}.

\section{Renormalization}
\label{sec4}

\subsection{Fixed points}

\begin{figure}
\begin{center}
\includegraphics[width=0.9 \columnwidth]{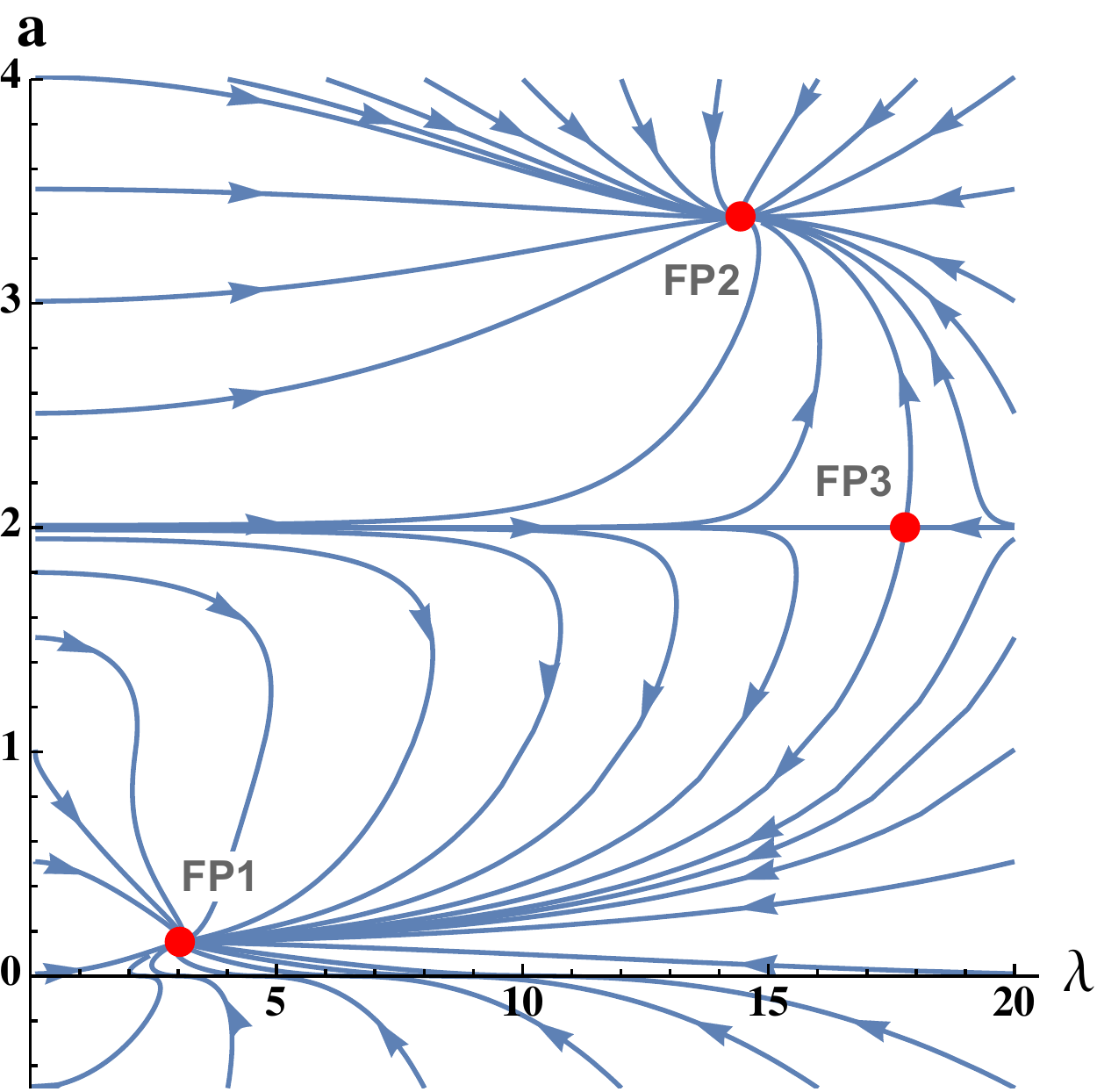}
\caption{RG flow in the $\lambda-a$ plane for $\epsilon=1/2$ and $N=2$. The three fixed points are marked by red dots and their coordinates are given in Eq.~\eqref{FPcoord}. We identify the stable fixed point FP1 with the $2k_F$ CDW quantum critical point of a metal with a generic Fermi surface, where two hot spots are connected by an incommensurate $2k_F$ wave vector.}
\label{fig2}
\end{center}
\end{figure}

We now use the minimal subtraction scheme to derive RG flow equations for all dimensionless parameters in Eq.~\eqref{Sfinal}. In order to make our theory UV finite we need to include counterterms in the action, which subtract the divergent terms in the limit $\epsilon \to 0$. These correspond to the $\sim1/\epsilon$ terms in the one-loop diagrams evaluated above. Since we used the convention $\overline{\Psi} \left( G_0^{-1} - \Sigma \right) \Psi$ for the definition of the self-energy $\Sigma$, we need to add the divergent part of self-energy to cancel the $1/\epsilon$-poles (the same holds for the bosonic self-energy $\Pi$). Therefore the renormalized action reads
	\beqn
		&S_{ren}& = S + S_{ct} \nn
		&=& \sum_j \int_{k^{d+1}} \overline{\Psi}_j(k) \left[ - i \bold{\Gamma} \cdot \K + i \sigma_x k_{d-1} Z_2 + i \sigma_x k_d^2 Z_3 \right] \Psi_j(k)\nn
		&+& \int_{k^{d+1}} \phi^+(k) \left[ k_d^2 Z_4 + a k_{d-1} Z_5 \right] \phi^-(-k) + S_{int},
	\eeqn
where we defined $Z_i = 1 + \frac{Z_{i,1}}{\epsilon}$ and
	\beqn
		&Z_{2,1}& = -\frac{2 u_1 \lambda^2}{(1-a) \sqrt{\vert \tilde{a} \vert} N},\ Z_{3,1} = \frac{2u_1 \lambda^2}{(1-a)^2 \sqrt{\vert \tilde{a} \vert} N},\nn
		&Z_{4,1}& = - \frac{u_1 \lambda^2}{2},\ Z_{5,1} = -\frac{u_1 \lambda^2}{a}.
	\eeqn
Introducing the rescaled bare fields 
	\beqn
		&&\K = \K_B,\ k_{d-1} = Z_2^{-1} k_{B,d-1},\ k_d = Z_3^{-\frac{1}{2}} k_{B,d},\nn
		 &&\Psi(k) = Z_2^{\frac{1}{2}} Z_3^{\frac{1}{4}} \Psi_B(k_B),\ \phi^{\pm}(k) = Z_2^{\frac{1}{2}} Z_3^{\frac{3}{4}} Z_4^{-1} \phi^{\pm}_B(k_B) \nn
		 &&\lambda_B = \lambda \mu^{\frac{\epsilon}{2}} Z_2^{-\frac{1}{2}} Z_3^{\frac{1}{4}} Z_4^{-\frac{1}{2}},\ a = Z_2 Z_3^{-1} Z_4 Z_5^{-1} a_B
		 \label{rescaling}
	\eeqn
brings the renormalized action back to its initial (bare) form in Eq.~\eqref{Sfinal}. The one-loop $\beta$-functions for the couplings $\lambda$ and $a$ follow straightforwardly from Eqs.~\eqref{rescaling} and take the form
	\beqn
		\beta_{\lambda} &=& \mu \frac{d\lambda}{d\mu} = \frac{u_1 \lambda^3}{2} \left(\frac{3-2a}{(1-a)^2 \sqrt{\vert \tilde{a} \vert} N} + \frac{1}{2} \right) - \frac{\epsilon}{2} \lambda, \nn
		\beta_a &=& \mu \frac{d a}{d \mu} = u_1 \lambda^2 \left(\frac{2a (2-a)}{(1-a)^2 \sqrt{\vert \tilde{a} \vert} N} + \frac{a}{2} -1 \right). 
	\eeqn
%Rewriting the arbitrary mass scale as $\mu = \mu_0 e^{-\ell}$ with a flow parameter $\ell$ and solving the resulting equations for fixed points, we find the three fixed points (for the physical case $N=2$)
For the physical case $N=2$ these $\beta$-functions describe three scale-invariant fixed points at
	\beqn
		&&\left( \lambda_1^*, a_1^* \right) = \left( 4.335 \sqrt{\epsilon}, 0.152 \right),\nn
		&&\left( \lambda_2^*, a_2^* \right) = \left( 20.43 \sqrt{\epsilon}, 3.383 \right),\nn
		&&\left( \lambda_3^*, a_3^* \right) = \left( 25.137 \sqrt{\epsilon}, 2.0 \right),
		\label{FPcoord}
	\eeqn
where the first and second fixed points are stable and the third one is unstable. The line $a=2$ separates the two domains of attraction of the two stable fixed points. A corresponding flow diagram is shown in Fig.~\ref{fig2}. Note that the $\beta$-functions are singular at $a=0,1$, but the differential equation for the RG flow trajectory $\frac{d a}{d\lambda}$ is regular at these points, giving rise to continuous solutions of the flow equations.

For the problem of interest here, namely a generic Fermi surface with two hot spots connected by a $2k_F$ wave vector, a physically sensible UV initial condition for the RG flow corresponds to a positive coupling $\lambda$ as well as an infinitesimally small value of $a$, such that the density susceptibility is initially peaked at the $\Q = \pm 2k_F$ wave vector. Consequently the RG flow is directed towards the first fixed point $\left( \lambda_1^*, a_1^* \right)$, which we identify with the continuous quantum phase transition between an ordinary Fermi liquid metal and the incommensurate $2k_F$ charge density wave phase. Note that an initial condition with $a=2$ would correspond to a perfectly circular Fermi surface, where the density susceptibility has degenerate maxima along the entire $2k_F$ line defined by $\pm 2 k_x + k_y^2 = 0$. In this highly fine tuned case fermions along the entire Fermi surface can scatter resonantly, not just at the two hot spots. It is important to realize that this doesn't invalidate the hot spot theory, however, because the scattering is local in momentum space and one obtains a theory with an infinite set of decoupled hot spot pairs. This situation is similar to the Ising-nematic problem and to the quantum phase transition between a normal metal and a FFLO superconductor at vanishing velocity detuning studied in Ref.~\cite{Pimenov}. Interestingly, the Fermi surface retains its shape and no dynamical nesting occurs during the flow along the $a=2$ line to the third fixed point $\left( \lambda_3^*, a_3^* \right)$. This is in stark contrast to the flow towards the first stable fixed point, where a strong dynamical nesting of the Fermi surface at the two hot spots occurs during the RG flow, as discussed in the next section. Finally, we do not identify the second stable fixed point $\left( \lambda_2^*, a_2^* \right)$ with a physically meaningful situation. A UV initial condition with an arbitrary value of $a$ different from zero or two would correspond to a density susceptibility with degenerate maxima that do not correspond to a $2k_F$ line.

\subsection{Correlators}

\begin{figure}
\begin{center}
\includegraphics[width=0.7 \columnwidth]{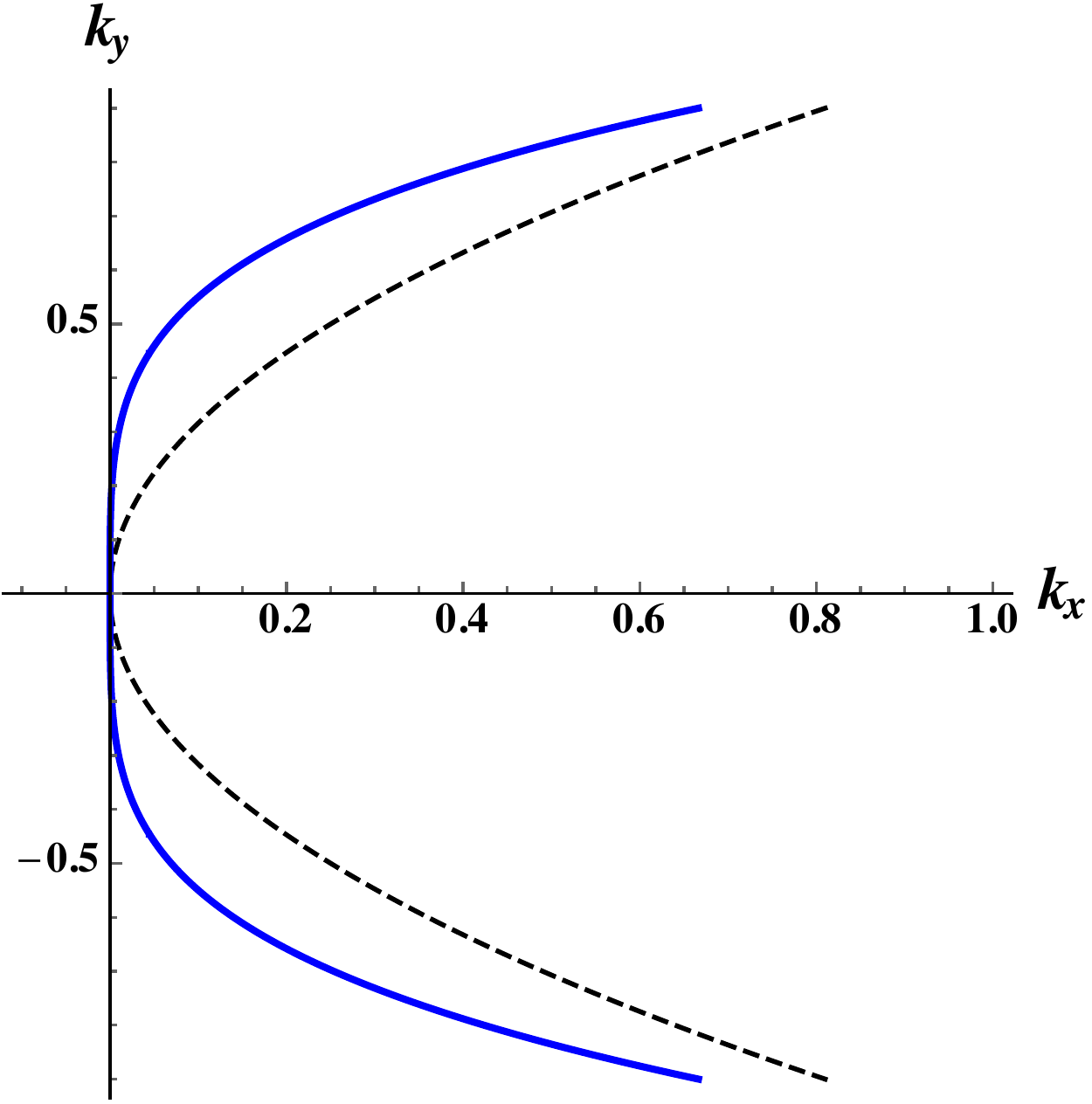}
\caption{Dynamical nesting of the Fermi surface. The blue solid line shows the form of the renormalized, flattened Fermi surface at one hot spot, as determined from Eq.~\eqref{FScondition} at the first RG fixed point $\left( \lambda_1^*, a_1^* \right)$ for $N=2$ and $\epsilon=1/2$, i.e.~in $d=2$ dimensions. The black dashed line indicates the initial, parabolic form of the Fermi surface as reference.}
\label{fig3}
\end{center}
\end{figure}

In the following we discuss the general scaling form of the correlation functions
	\beqn
		&&\langle \Psi(k_1) ... \Psi(k_m) \overline{\Psi}(k_{m+1}) ... \overline{\Psi}(k_{2m}) \nn
		&&\phi^+(k_{2m+1}) ... \phi^+(k_{2m+n}) \phi^-(k_{2m+n+1}) ... \phi^-(k_{2m+2n})\rangle \nn
		&&= G^{(m,m,n,n)}\left( \{ k_i \}, \mu, \lambda, a \right) \delta^{(d+1)}\left( \{ k_i \} \right),
	\eeqn
where $\delta^{(d+1)}\left( \{ k_i \} \right)$ ensures energy and momentum conservation. We define the dynamical critical exponents
	\beqn
		&&z_{d-1}^{-1} = 1+ \frac{d \ln Z_2}{d \ln \mu} = 1+ \frac{2 u_1 \lambda^2}{(1-a) \sqrt{\vert \tilde{a} \vert} N}, \nn
		&&z_d^{-1} = 1 + \frac{d \ln Z_3}{d \ln \mu} = 1- \frac{2 u_1 \lambda^2}{(1-a)^2 \sqrt{\vert \tilde{a} \vert} N},
		\label{dyncritexp}
	\eeqn
as well as the anomalous dimensions of the fermion and boson fields
	\beqn
		&&\eta_{\Psi} = \frac{1}{2} \frac{d \ln Z_{\Psi}}{d \ln \mu} = \frac{u_1 \lambda^2}{2} \frac{2a-1}{(1-a)^2 \sqrt{\vert \tilde{a} \vert} N}, \nn
		&&\eta_{\phi} = \frac{1}{2} \frac{d \ln Z_{\phi}}{d \ln \mu} = \frac{u_1 \lambda^2}{2} \left( \frac{2a+1}{(1-a)^2 \sqrt{\vert \tilde{a} \vert}N} + \frac{1}{2} \right),
	\eeqn
where $Z_{\Psi} = Z_2^{-1} Z_3^{-1/2}$ and $Z_{\phi} = Z_2^{-1} Z_3^{-3/2} Z_4$. Using these quantities the renormalization group equation for the correlation functions takes the form
	\beqn
		&&\Bigg[ \sum_{i=1}^{2m+2n} \left( \K_i \nabla_{\K_i} + \frac{k_{d-1,i}}{z_{d-1}} \frac{\partial}{\partial k_{d-1,i}} +  \frac{k_{d,i}}{2 z_d} \frac{\partial}{\partial k_{d,i}} \right) \nn
		&&-\beta_{\lambda} \frac{\partial}{\partial \lambda} -\beta_{a} \frac{\partial}{\partial a} - 2m \left( \eta_{\Psi} - \frac{4-\epsilon}{2}\right) - 2n \left( \eta_{\phi} - \frac{4-\epsilon}{2} \right) \nn
		&&+ \left( \epsilon - \frac{3}{2} - \frac{1}{z_{d-1}} - \frac{1}{2 z_d} \right) \Bigg] G^{(m,m,n,n)}\left( \{ k_i \}, \mu, \lambda, a \right) =0.\nn
	\eeqn
At the fixed points, where the $\beta$-functions are zero, the solution of the RG equation for the fermion and boson two-point functions gives rise to the scaling forms
%	\beqn
%		G(k) &=& \frac{\text{sgn}(k_{d-1})}{\vert k_{d-1} \vert^{z_{d-1}}} f_\Psi \! \left( \frac{\vert \K \vert}{c(k_{d-1})}, \frac{\vert k_d \vert^{2 z_d}}{c(k_{d-1})} \right), \\
%		D_+(k) &=& \frac{\text{sgn}(k_{d-1})}{\vert k_{d-1} \vert^{p z_{d-1}}} f_\phi \! \left( \frac{\vert \K \vert}{c(k_{d-1})}, \frac{\vert k_d \vert^{2 z_d}}{c(k_{d-1})} \right), \label{Dscale}
%	\eeqn
%
	\beqn
		G(k) &=& \frac{1}{\vert k_d \vert^{2 z_d}} f_\Psi \! \left( \frac{\vert \K \vert}{\vert k_d \vert^{2 z_d}}, \frac{ \text{sgn}(k_{d-1}) \vert k_{d-1} \vert^{z_{d-1}} }{\vert k_d \vert^{2 z_d}} \right) \! , \ \ \ \ \ \label{Gscale} \\
		D_+(k) &=& \frac{1}{\vert k_d \vert^{2 p z_d}} f_\phi \! \left( \frac{\vert \K \vert}{\vert k_d \vert^{2 z_d}}, \frac{ \text{sgn}(k_{d-1}) \vert k_{d-1} \vert^{z_{d-1}} }{\vert k_d \vert^{2 z_d}} \right) \! , \label{Dscale}
	\eeqn
with
	\beq
%		c(k_{d-1}) = \text{sgn}(k_{d-1}) \vert k_{d-1} \vert^{z_{d-1}}, \ \
		p = \frac{1}{z_d} - \frac{u_1 (\lambda^*)^2}{2}
	\eeq
and $f_\Psi$ and $f_\phi$ are universal scaling functions. From the scaling form of the fermion propagator we can infer the shape of the renormalized Fermi surface at the hot spots. In the non-interacting case the fermion propagator has poles at the Fermi surface defined by $f^{-1}_\Psi(0,\pm 1)=0$. Analogously, the renormalized shape of the Fermi surface is then determined by the equation
	\beq
		\text{sgn}(k_{d-1}) \vert k_{d-1} \vert^{z_{d-1}} = \pm \vert k_d \vert^{2 z_d} .
		\label{FScondition}
	\eeq
%For $\epsilon = \frac{1}{2}$, $N=2$ the critical exponents take the following values at the three fixed points:
%	\beqn
%		(z_{d-1}, z_d)_1^* &=& \left( 1.283, \frac{2}{3} \right), \nn
%		(z_{d-1}, z_d)_2^* &=& \left( 0.206, \frac{2}{3} \right), \nn
%		(z_{d-1}, z_d)_3^* &=& \left( - \sqrt{2}, -\sqrt{2} \right).
%	\eeqn
For the physical case $N=2$ the first fixed point $\left( \lambda_1^*, a_1^* \right)$ is characterized by the dynamical critical exponents
	\beq
		(z_{d-1}^{-1}, z_d^{-1})_1^* = \left( 1+0.566 \epsilon, 1-\frac{2}{3} \epsilon \right) 
	\eeq
and anomalous dimensions of the fermion and boson fields given by 
	\beq
		(\eta_\Psi, \eta_\phi)_1^* = \left( -0.116 \epsilon , 0.268 \epsilon \right).
	\eeq
At this fixed point the Fermi surface in the vicinity of the two hot spots takes the form $k_x = \pm \vert k_y \vert ^{3.85}$ in $d=2$ dimensions, which indicates a strong dynamical nesting with a vanishing Fermi surface curvature at the hot spots, as shown in Fig.~\ref{fig3}. By contrast, for $a=2$ the two critical exponents $z_d$ and $z_{d-1}$ in Eq.~\eqref{dyncritexp} are equal and thus the Fermi surface retains its shape for the RG flow along the $a=2$ line. Note that all fixed points describe non-Fermi liquids, with a non-linear, power-law fermion dispersion perpendicular (tangential) to the Fermi surface determined by the dynamical critical exponent $z_{d-1}$ ($2 z_d$).

\section{Experimental Signatures}
\label{sec5}

While there ist strong evidence that the incommensurate CDW transition in rare-earth tellurides is driven by Fermi surface nesting \cite{Gweon, Yao}, the situation in the transition metal dichalcogenides NbSe$_2$ and TaS$_2$ is much less clear. Fermi surface nesting \cite{Feng, Straub}, saddle bands \cite{Rice}, as well as electron-phonon coupling \cite{Weber, Flicker} have been suggested as possible CDW mechanisms. Here we argue that it would be beneficial to study these materials in the vicinity of the quantum phase transition to the density wave ordered phase. Experimentally, the quantum phase transition can be driven by applying pressure \cite{Feng, Feng2} or intercalating different transition metals \cite{Morosan}. As we've layed out in this work, a quantum phase transition driven by Fermi surface nesting has definite signatures, which can be detected in the quantum critical regime at finite temperature above the quantum critical point. The most striking experimental signature of CDW quantum criticality would be the power-law behavior of the density susceptibility in the vicinity of the critical point. At the $2k_F$ wave vector $\Q$ its characteristic power-law frequency dependence follows from Eq.~\eqref{Dscale} and has the form $D(\omega) \sim \vert \omega \vert^{-p}$ with the exponent $p=1-0.769 \epsilon \simeq 0.616$ for the physical case $N=2$ and in $d=2$ dimensions, which should be observable with a variety of experimental probes, such as Raman- or inelastic X-ray and neutron scattering. The same signatures should be observable in rare-earth tellurides such as SmTe$_3$ and TbTe$_3$.

Moreover, in the quantum critical regime we expect $\omega/T$ scaling, i.e.~the temperature $T$ and frequency $\omega$ dependent density susceptibility at the $2k_F$ wave vector should obey the scaling relation
	\beq 
		D(\omega,T) = \vert \omega \vert^{-p} \, f_T(T/\omega)
	\eeq
 with a universal scaling function $f_T$. Accordingly, the static density susceptibility at the $2k_F$ wave vector has a power-law temperature dependence $D(0) \sim T^{-p}$ with the same exponent $p$. Again, this characteristic temperature dependence should be observable using Raman- or elastic X-ray and neutron scattering.

Thermodynamic signatures are a bit more difficult to discern, unfortunately. The specific heat has singular contributions from ''hot'' fermions in the vicinity of the hot spots, as well as an ordinary linear in temperature contribution from ''cold'' fermions far away from the hot spots. The latter contribution dominates, however, which can be seen as follows: in analogy to the spin density wave critical point we expect hyperscaling to be obeyed in the CDW hot spot theory discussed here \cite{PatelStrack}, consequently the scaling dimension of the free energy density is given by $[f]= (d-1) + z_{d-1}^{-1} + (2 z_d)^{-1}$, where the first term comes from the scaling dimension of the time-like- and the extra-dimensional directions, whereas the last two terms arise from the scaling dimension of the two spatial directions. The temperature dependence of the specific heat thus takes the form 
	\beq
		C_v \sim T^{d-2+z_{d-1}^{-1} + (2 z_d)^{-1}} \ .
	\eeq
 In two spatial dimensions the specific heat exponent is larger than one at the first fixed point and thus the singular contribution is subleading compared to the $C_v \sim T$ contribution from cold fermions.

An alternative way to compute this contribution is via the scaling form of the electron propagator in Eq.~\eqref{Gscale}, from which the electronic density of states $\nu(\omega)$ of the hot electrons can be computed. In $d=2$ spatial dimensions it takes the form 
	\beq
		\nu(\omega) - \nu(0) \sim (1-z_{d-1}) \, |\omega|^{z_{d-1}^{-1}+(2 z_d)^{-1}-1} \ . 
	\eeq 
The singular part of $\nu(\omega)$ gives rise to the same temperature dependence of $C_v$ as determined above. Note, however, that the non-zero constant density of states $\nu(0)$ again leads to a dominant linear contribution $C_v \sim T$.

\section{Superconducting instabilities}
\label{sec6}

Superconductivity has been observed in the vicinity of various metallic quantum critical points. For the $2k_F$ CDW critical point discussed here, a natural superconducting instability corresponds to the formation of spin-singlet Cooper pairs between electrons at the two antipodal hot spots. In order to investigate whether such a superconducting instability is enhanced or suppressed in the vicinity of the QCP, we compute the scaling dimension of the singlet Cooper-pair creation operator at the fixed point by including a corresponding source term in the action:
	\beqn
		S_{cp} = g \int_{k} \Big[ &\psi&_{+,\uparrow}(k) \psi_{-,\downarrow}(-k) \nn
		- &\psi&_{+,\downarrow}(k) \psi_{-,\uparrow}(-k) + c.c. \Big].
	\eeqn
In spinor representation and in general dimensions this term can be written as
	\beq
		S_{cp} = g \int_{k^{d+1}} \tau_y^{\alpha \beta} \overline{\Psi}_{\alpha}(k) \Psi_{\beta}(k),
	\eeq
where the Greek indices are spin indices and $\tau_y^{\alpha \beta}$ is a Pauli matrix in spin space. The vertex factor and the one-loop correction read
$-g \tau_y^{\alpha \beta} \mathbb{1} +g \tau_y^{\alpha \beta} V$,
where $V$ is a matrix in spinor space given by the integral
	\beq
		V = \frac{\lambda^2 \mu^{\epsilon}}{N} \int_p \sigma_y G^2(p) \sigma_y D_+(p+q). \label{supinst}
	\eeq
Calculating $V$ (see Appendix \ref{appendixsi}) leads to the $1/\epsilon$-pole
	\beq
		V = - \mathbb{1} \frac{4 u_1 \lambda^2}{(1-a) \sqrt{\vert \tilde{a} \vert} N \epsilon} + \tn{finite terms} \label{supinstpole}
	\eeq
The renormalized action, which we get by adding $V$, is given by
	\beq
		S_{cp}^r = g \mu Z_g \int_{k^{d+1}} \tau_y^{\alpha \beta} \overline{\Psi}_{\alpha}(k) \Psi_{\beta}(k),
	\eeq
where we introduced an arbitrary mass scale $\mu$ to make the coupling dimensionless and $Z_g = 1 + \frac{Z_{g,1}}{\epsilon}$ with $Z_{g,1} = - \frac{4 u_1 \lambda^2}{(1-a) \sqrt{\vert \tilde{a} \vert} N}$. The $\beta$-function for the source field $g$ reads
	\beq
		\beta_g = g (-1-\eta_g)
	\eeq
with the anomalous dimension
	\beq
		 \eta_g = \frac{d \ln Z_g}{d \ln \mu} = \frac{4 u_1 \lambda^2}{(1-a) \sqrt{\vert \tilde{a} \vert} N}.
	\eeq
The first fixed point is the only one with $a<1$ and therefore $\eta_g = 1.131 \epsilon > 0$ for $N=2$. Consequently superconducting instabilities are enhanced at the first fixed point. At the other two fixed points $\eta_g<0$ and superconductivity is suppressed.

\section{Discussion and Conclusions}
\label{sec7}

We've presented a controlled, perturbative renormalization group study of incommensurate $2k_F$ CDW quantum critical points in two-dimensional metals, which treats electronic as well as bosonic order parameter fluctuations on equal footing, and found a stable fixed point corresponding to a continuous quantum phase transition with a strongly renormalized, flattened Fermi surface at the hot spots. This result is in contrast to the early theoretical observation of a first order incommensurate transition due to strong fluctuations in Ref.~\cite{Altshuler}, which did not take the crucial dynamical nesting of the Fermi surface into account, however. Indeed, Sykora \emph{et al.}~pointed out in Ref.~\cite{Sykora} that the Fermi surface is strongly renormalized and flattened at the two hot spots connected by the $2k_F$ wave vector and our results strengthen their observation. Moreover, we've presented experimentally testable predictions for the density susceptibility in the vicinity of the quantum critical point, which could help to clarify the controversially debated CDW mechanism in the transition metal dichalcogenides NbSe$_2$ and TaS$_2$. 

One important point that we haven't discussed so far is how to tune our model away from criticality by adding a boson mass term to the action. Naively, adding a boson mass in Eq.~\eqref{Sfinal} would turn the boson massless at a momentum different from $k=0$ rather than leading to a gap, due to the linear term $\sim a k_{d-1}$ in the boson propagator. This seemingly invalidates our assumption that the density susceptibility is peaked at the $2k_F$ momentum $\Q$, i.e.~at $k=0$. It is crucial to realize, however, that the boson propagator strongly renormalizes during the RG flow and the dynamical nesting of the Fermi surface ensures that the boson self-energy remains peaked at $k=0$. Perturbing the fixed point action with a mass term thus gaps out the boson at all momenta. A simple way to see this is as follows:
the detailed form of the one-loop boson self-energy in Eqs.~\eqref{bosone>0} and \eqref{bosone<0} computed with the bare fermion propagators shows that it is peaked at $k=0$ in dimensions $d<2$, but doesn't have a peak at $k=0$ for $d \geq 2$. This behavior is reminescent of the Lindhard density susceptibility of a free Fermi gas. By contrast, computing the boson self-energy using a renormalized, nested fermion dispersion with vanishing Fermi surface curvature at the two hot spots of the form $\pm k_{d-1}+|k_d|^\alpha$ with $\alpha > 2$ leads to a well defined peak in the $d=2$ density susceptibility at the $2k_F$ wave vector $\Q$, i.e.~at $k=0$.

Another technical detail worthwile to discuss is that our computation crucially differs from an analogous approach to Ising-nematic quantum criticality in one aspect: while Ref.~\cite{Dalidovich} had to reorganize the perturbation expansion by including the Landau damping term in the boson propagator from the start to remove IR divergences, this is not necessary for the CDW problem considered here, where similar IR divergences are cured by the linear $\sim k_{d-1}$ term in the boson propagator. As a consequence, Landau damping effects are subleading and only appear at two-loop order in the problem considered here. We leave the challenging computation of two-loop effects open for future investigation. 

\acknowledgements

We thank W. Metzner and J. Sykora for helpful discussions. This work is supported by the Deutsche Forschungsgemeinschaft (DFG, German Research Foundation) under Germany's Excellence Strategy via the Nanosystems-Initiative Munich (NIM), as well as the Munich Center for Quantum Science and Technology (MCQST) - EXC-2111 - 390814868.

\appendix

\section{Boson self-energy}
\label{appA}

The integral for the one-loop boson self-energy in Eq.~\eqref{bosonselfenergy} can be evaluated as follows: using the properties of the Pauli matrices, the trace can be simplified using
	\beq
		\tn{Tr}\left( \sigma_y \sigma_i \sigma_y \sigma_j \right) = \begin{cases} 2 & i=j=y \\ -2 & i=j\neq y \\ 0 & i \neq j
		\end{cases},
	\eeq
which yields
	\beq
		\Pi(k) = - \lambda^2 \mu^{\epsilon} \int_p \frac{\delta_p \delta_{k-p} - \P \cdot (\P - \K)}{\left( \P^2 + \delta_p^2 \right) \big( (\P-\K)^2 + \delta_{k-p}^2 \big)}.
	\eeq
Shifting $p_{d-1} \rightarrow p_{d-1}-p_d^2$ and introducing the new integration variable $y = \frac{1}{\sqrt{2}} (2p_d-k_d)$ as well as $e_k = k_{d-1} + \frac{k_d^2}{2}$ leads to
	\begin{widetext}
	\beq
		\Pi(k) = \frac{\lambda^2 \mu^{\epsilon}}{\sqrt{2}} 
		\int_{\P,p_{d-1},y} \frac{p_{d-1} (p_{d-1} -e_k-y^2) + \P \cdot (\P-\K)}{\left( \P^2 + p_{d-1}^2 \right) \left( (\P-\K)^2 + (p_{d-1} - e_k - y^2)^2 \right)}.
	\eeq
Using the Feynman parametrization
%	\beq
%		\frac{1}{A B} = \int_0^1 dx \frac{1}{\big[ xA+(1-x)B \big]^2}
%	\eeq
%and $\P \rightarrow \P + x\Q$ 
the above integral can be written as
	\beq
		\Pi(k) = \frac{\lambda^2 \mu^{\epsilon}}{\sqrt{2}} \int_0^1dx \int \frac{d^{d-1}\P dp_{d-1} dy}{(2\pi)^{d+1}} \frac{p_{d-1}^2-(e_k+y^2)p_{d-1} + \P^2 - x(1-x) \K^2}{\Big[ p_{d-1}^2 -2x(e_k+y^2) p_{d-1} + x(e_k+y^2)^2 + \P^2 + x(1-x) \K^2 \Big]^2}.
	\eeq
Integrating over $p_{d-1}$ and then rescaling $\P \rightarrow \sqrt{x(1-x)} \P$ yields
	\beq
		\Pi(q) = \frac{\lambda^2 \mu^{\epsilon}}{\sqrt{8}} \int_0^1dx\ [x(1-x)]^{\frac{d}{2} - 1} \int \frac{d^{d-1}\P dy}{(2\pi)^{d}} \frac{\P^2}{\big[ \P^2 + \K^2 + (e_k+y^2)^2 \big]^{\frac{3}{2}}}.
	\eeq
	\end{widetext}
The $x$-integral is elementary and after switching to hyperspherical coordinates, substracting $\Pi(0)$ for UV regularization and performing the integral over the radial coordinate we obtain
%	\beqn
%	&&\Pi(q)-\Pi(0) = \lambda^2 \mu^{\epsilon} \frac{\Gamma^2\left( \frac{d}{2} \right)}{2^{d-\frac{1}{2}} \pi^{\frac{d+1}{2}} \eulergamma{d} \eulergamma{\frac{d-1}{2}}}  \times \nn && 
%	\int_0^{\infty} \! dy dr \! \left( \frac{r^d}{\big[ r^2 + \K^2 + (e_k+y^2)^2 \big]^{\frac{3}{2}}} - \frac{r^d}{\big[ r^2 + y^4 \big]^{\frac{3}{2}}} \right). \nn
%	\eeqn
%The integration over $r$ is then straightforward and leads to
	\beqn
		&&\Pi(q) - \Pi(0)=\lambda^2 \mu^{\epsilon} \frac{\Gamma^2\left( \frac{d}{2} \right) \eulergamma{\frac{d+1}{2}} \eulergamma{1-\frac{d}{2}}}{2^{d-\frac{1}{2}} \pi^{\frac{d}{2} +1} \eulergamma{d} \eulergamma{\frac{d-1}{2}}}  \times \nn
		&& \int_0^{\infty}dy \left[ \left( \K^2 + (e_k+y^2)^2 \right)^{\frac{d}{2}-1}- (y^4)^{\frac{d}{2}-1} \right].
	\eeqn
Since the integral has different solutions for $e_k>0$ and $e_k<0$ we need to distinguish the two cases. After setting $d= \frac{5}{2}-\epsilon$, the self-energy for $e_k>0$ reads 
	\begin{widetext}
	\beq
		\Pi(q) - \Pi(0) = \lambda^2 \mu^{\epsilon} \frac{\Gamma^2\left( \frac{5}{4}- \frac{\epsilon}{2} \right) \eulergamma{-\frac{1}{4}+\frac{\epsilon}{2}} \eulergamma{-1+\epsilon}}{2^{4-\epsilon} \pi^{\frac{7}{4} - \frac{\epsilon}{2}} \eulergamma{\frac{5}{2}-\epsilon} \eulergamma{-\frac{1}{2} +\epsilon}}  \left( \frac{3}{2}-\epsilon \right) e_k^{1-\epsilon} \, {}_2F_1\left[ \frac{\epsilon-1}{2}, \frac{\epsilon}{2}, \frac{2\epsilon+1}{4}, - \frac{\K^2}{e_k^2} \right], \label{bosone>0}
	\eeq
which gives the $1/\epsilon$-pole of Eq.~(\ref{bosonpole}) when expanding the hypergeometric function ${}_2F_1$ around $\epsilon=0$. 
In the case of $e_k<0$, the solution to the $y$-integration is 

	\beqn
		\Pi(q)-\Pi(0) = \lambda^2 \mu^{\epsilon} \frac{\Gamma^2\left( \frac{5}{4}- \frac{\epsilon}{2} \right) \left( \frac{3}{2}-\epsilon \right)}{2^{4-\epsilon} \pi^{\frac{9}{4} - \frac{\epsilon}{2}} \eulergamma{\frac{5}{2}-\epsilon}} \vert \K \vert^{-\epsilon} \Bigg(&& 2 \vert \K \vert \eulergamma{\frac{5}{4}} \eulergamma{\frac{\epsilon-1}{2}} {}_2F_1\left[ \frac{1}{4}, \frac{\epsilon-1}{2}, \frac{1}{2}, - \frac{e_k^2}{\K^2} \right]\nn
		&&- e_k \eulergamma{\frac{3}{4}} \eulergamma{\frac{\epsilon}{2}} {}_2F_1\left[ \frac{3}{4}, \frac{\epsilon}{2}, \frac{3}{2}, - \frac{e_k^2}{\K^2} \right] \Bigg), 
		\label{bosone<0}
	\eeqn
	\end{widetext}
which leads to the same $1/\epsilon$ pole as the solution for $e_k>0$ above. Expanding this expression in $\vert \K \vert$ and afterwards around $\epsilon = 0$ yields the expression in (\ref{bosonpole}). Note that doing the same with (\ref{bosone>0}), the lowest order term in $\vert \K \vert$ is $\propto \K^2$.

\section{Fermion self-energy}
\label{appB}

The product of Pauli matrices in Eq.~\eqref{fermionselfenergy} can be simplified to
	\beqn
		\sigma_y G^T(k) \sigma_y &&= -i \sigma_y \frac{k_0 \sigma_y - \k \cdot \sigma_z + \delta_k \sigma_x}{\K^2+ \delta_k^2} \sigma_y\nn
		&&= -i \frac{k_0 \sigma_y + \k \cdot \sigma_z - \delta_k \sigma_x}{\K^2+ \delta_k^2} = -G(k)
	\eeqn
and therefore
	\beqn
		&&\Sigma(k) = - \frac{\lambda^2 \mu^{\epsilon}}{N} \int_p G(p-k) D_+(p) \nn
		&&= \frac{i \lambda^2 \mu^{\epsilon}}{N} \int_p \frac{- \mathbf{\Gamma} \cdot (\P-\K) + \sigma_x \delta_{p-k}}{(\P-\K)^2 + \delta^2_{p-k}} \frac{1}{p_d^2 + ap_{d-1}}.
	\eeqn
Shifting $\P \rightarrow \P+\K$, $p_{d-1} \rightarrow p_{d-1}-p_d^2 +2k_dp_d$ and defining $\tilde{a} = \frac{a}{1-a}$ leads to
	\beqn
		\Sigma(k) = \frac{i \lambda^2 \mu^{\epsilon} \sigma_x}{(1-a)N} \int_p &&\frac{p_{d-1} + \delta_{-k}}{\P^2 + (p_{d-1} + \delta_{-k})^2}\times \nn
		&&\frac{1}{p_d^2 + 2 \tilde{a} k_d p_d + \tilde{a} p_{d-1}},
	\eeqn
where $\delta_{-k} = - k_{d-1} + k_d^2$. The $p_d$-integral can be evaluated using the principal value, which leads to
	\beqn
		\Sigma(k) = \frac{i \lambda^2 \mu^{\epsilon} \sigma_x}{2(1-a) \sqrt{\vert \tilde{a} \vert} N} &&\int \frac{d^{d-1}\P dp_{d-1}}{(2\pi)^d} \frac{p_{d-1} + \delta_{-k}}{\P^2 + (p_{d-1} + \delta_{-k})^2} \nn
		&& \frac{\Theta\left( \tn{sgn}(\tilde{a}) p_{d-1} - \vert \tilde{a} \vert k_d^2 \right)}{\sqrt{\tn{sgn}(\tilde{a}) p_{d-1} - \vert \tilde{a} \vert k_d^2}}.
	\eeqn
After the substitution $y= \tn{sgn}(\tilde{a}) p_{d-1} - \vert \tilde{a} \vert k_d^2$, the fermion self-energy reads
	\beqn
		\Sigma(q) = &&\frac{i \lambda^2 \mu^{\epsilon} \sigma_x \tn{sgn}(\tilde{a})}{2(1-a) \sqrt{\vert \tilde{a} \vert} N} \int\frac{d^{d-1}\P}{(2\pi)^{d-1}} \int_0^{\infty}\frac{dy}{2\pi} \nn
		&&\frac{y+c(k)}{\P^2 + \big(y+c(k) \big)^2} \frac{1}{\sqrt{y}},
	\eeqn
where we defined $c(k) = \vert \tilde{a} \vert k_d^2 + \tn{sgn}(\tilde{a}) \delta_{-k}$. The $y$-integration is elementary and using hyperspherical coordinates we get
	\beqn
		&&\Sigma(k) - \Sigma(0) =\frac{i \lambda^2 \mu^{\epsilon} \sigma_x \tn{sgn}(\tilde{a})}{(1-a) \sqrt{\vert \tilde{a} \vert} N} \frac{1}{2^d \pi^{\frac{d-1}{2}} \eulergamma{\frac{d-1}{2}}} \times \nn
		&& \tn{Re} \left\{ \int_0^{\infty}dr\ r^{d-2} \left( \frac{1}{\sqrt{c(k) + ir}} - \frac{1}{\sqrt{ir}} \right) \right\},
	\eeqn
where we substracted $\Sigma(0)$ for UV-regularization since the above integral initially converges only for $d<\frac{3}{2}$. Carrying out the $r$-integral gives the result
	\beqn
		\Sigma(k)-\Sigma(0) =&&\frac{i \lambda^2 \mu^{\epsilon} \sigma_x \tn{sgn}(\tilde{a})}{(1-a) \sqrt{\vert \tilde{a} \vert} N} \frac{\eulergamma{\frac{3}{2}-d} \eulergamma{d-1}}{2^d \pi^{\frac{d}{2}} \eulergamma{\frac{d-1}{2}}} \times \nn
		&& \tn{Re}\left\{ i^{1-d} \big( c(k) \big)^{d-\frac{3}{2}} \right\}.
	\eeqn
Setting $d= \frac{5}{2}-\epsilon$ and expanding around $\epsilon = 0$ gives the pole contribution in Eq.~(\ref{fermionpole}).

\section{Superconducting vertex}
\label{appendixsi}

The matrix product in the integral (\ref{supinst}) can be evaluated to
	\beqn
		\sigma_y G^2(p) \sigma_y &=& - \mathbb{1} \frac{p_0^2 + \sum_{i,j=1}^{d-2} p_i p_j + \delta_p^2}{\left( \P^2 + \delta_p^2 \right)^2}\nn
		&=& - \mathbb{1} \frac{p_0^2 + \p^2 + \sum_{i \neq j} p_i p_j + \delta_p^2}{\left( \P^2 + \delta_p^2 \right)^2}.
	\eeqn
The term $\sum_{i \neq j} p_i p_j$ vanishes by antisymmetry under $p_i \rightarrow -p_i$ since the boson propagator is independent of $p_i$ and so we get after shifting $p_{d-1} \rightarrow p_{d-1}-p_d^2$
	\beqn
		V = &&- \mathbb{1} \frac{\lambda^2 \mu^{\epsilon}}{(1-a)N} \int_p \frac{1}{\P^2+p_{d-1}^2} \times \nn
		&&\frac{1}{p_d^2 + \frac{2q_d}{1-a} p_d + \frac{1}{1-a} q_d^2 + \tilde{a} p_{d-1} + \tilde{a} q_{d-1}}.
	\eeqn
Here the $p_d$-integral is straightforward using the principal value and with the substitution $y = \tn{sgn}(\tilde{a}) p_{d-1}-f(q)$ where $f(q) = -\tn{sgn}(\tilde{a})p_{d-1}+\tn{sgn}(\tilde{a}) \frac{1}{1-a}q_d^2$ the vertex correction reads
	\beqn
		V = &&- \mathbb{1} \frac{\lambda^2 \mu^{\epsilon}}{2(1-a) \sqrt{\vert \tilde{a} \vert} N} \int \frac{d^{d-1}\P}{(2\pi)^{d-1}} \int_0^{\infty} \frac{dy}{2\pi}\nn
		&&\frac{1}{\P^2 + \big( y+f(q) \big)^2} \frac{1}{\sqrt{y}}.
	\eeqn
Integrating over $y$ and changing to hyperspherical coordinates yields
	\beqn
		V = &&-\mathbb{1} \frac{\lambda^2 \mu^{\epsilon}}{(1-a) \sqrt{\vert \tilde{a} \vert} N} \frac{i}{2^{d+1} \pi^{\frac{d-1}{2}} \eulergamma{\frac{d-1}{2}}} \int_0^{\infty}dr \times \nn
		&& r^{d-3} \left( \frac{1}{\sqrt{f(q)+ir}} - \frac{1}{\sqrt{f(q)-ir}} \right),
	\eeqn
which can be evaluated to
	\beqn
		V = &&\mathbb{1} \frac{\lambda^2 \mu^{\epsilon}}{(1-a) \sqrt{\vert \tilde{a} \vert} N} \frac{i\ \eulergamma{\frac{5}{2} -d} \eulergamma{d-2} \big( f(q) \big)^{\frac{5}{2}-d}}{2^{d+1} \pi^{\frac{d}{2}} \eulergamma{\frac{d-1}{2}}} \nn
		&& \big[ i^{-d} - (-i)^{-d} \big].
	\eeqn
Setting $d= \frac{5}{2}-\epsilon$ and expanding around $\epsilon=0$ leads to the expression in (\ref{supinstpole}).

\end{document}